\documentclass{article}
\usepackage{xspace}

\PassOptionsToPackage{numbers,compress}{natbib}
\usepackage[final]{neurips}

\usepackage{latexsym}
\usepackage{tcolorbox}

\usepackage[T1]{fontenc}
\usepackage[utf8]{inputenc}
\usepackage{microtype}

\usepackage{inconsolata}
\usepackage{wrapfig,lipsum,booktabs}
\usepackage{pifont}
\usepackage{makecell} 
\usepackage{tabularx}
\usepackage[normalem]{ulem}
\useunder{\uline}{\ul}{}
\usepackage[export]{adjustbox}

\usepackage{hyperref}       
\usepackage{url}            
\usepackage{booktabs}       
\usepackage{amsfonts}       
\usepackage{nicefrac}       

\usepackage{graphicx}
\usepackage{subfigure}
\usepackage{float}
\usepackage{multirow}

\usepackage{listings}

\usepackage{amsmath}
\usepackage{amssymb}
\usepackage{mathtools}
\usepackage{amsthm}
\usepackage{balance}
\usepackage{flushend}

\usepackage{lipsum}
\usepackage{graphicx}
\usepackage{subcaption}
\usepackage{tikz}
\usetikzlibrary{shapes, arrows, positioning}

\usepackage{fancyhdr}

\hypersetup{
	colorlinks,
	citecolor=gray,
	linkcolor=red,
	urlcolor=blue}

\usepackage[capitalize,noabbrev]{cleveref}
\usepackage{tablefootnote}

\usepackage{color, colortbl,xcolor}
\usepackage{enumitem}
\usepackage{booktabs,arydshln}

\newcommand{\nbf}[1]{{\noindent \textbf{#1}}}

\newcommand{\gene}{\texttt{Gene}\xspace}
\newcommand{\skill}{\texttt{Skill}\xspace}
\newcommand{\evolver}{\texttt{Evolver}\xspace}

\makeatletter
\def\adl@drawiv#1#2#3{%
	\hskip.5\tabcolsep
	\xleaders#3{#2.5\@tempdimb #1{1}#2.5\@tempdimb}%
	#2\z@ plus1fil minus1fil\relax
	\hskip.5\tabcolsep}
\newcommand{\cdashlinelr}[1]{%
	\noalign{\vskip\aboverulesep
		\global\let\@dashdrawstore\adl@draw
		\global\let\adl@draw\adl@drawiv}
	\cdashline{#1}
	\noalign{\global\let\adl@draw\@dashdrawstore
		\vskip\belowrulesep}}
\makeatother

\setlist[itemize]{align=parleft,left=0pt..0.5em}
\setlist[enumerate]{align=parleft,left=0pt..0.5em}

\setlist[itemize]{align=parleft,left=0pt..0.8em}

\newcommand{\evomapheader}{%
  \noindent
  \begin{minipage}[c]{0.62\linewidth}
    \href{https://evomap.ai}{%
      \includegraphics[
        height=14pt,
        keepaspectratio
      ]{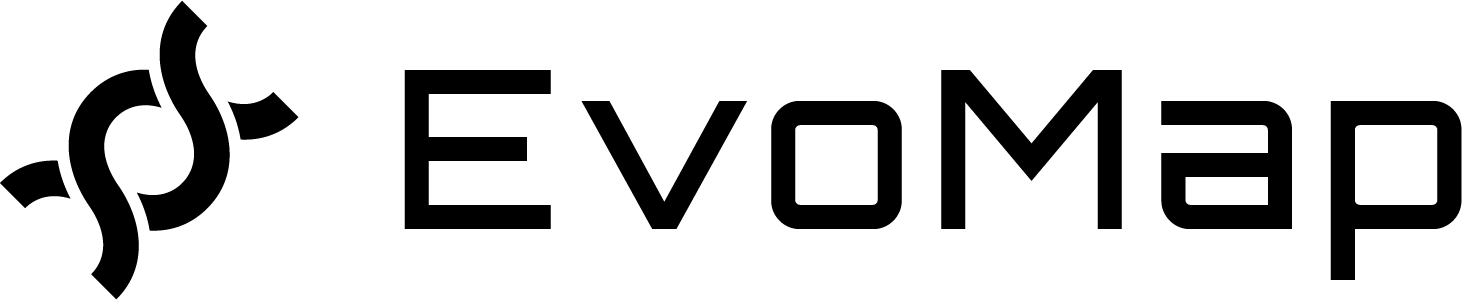}}
  \end{minipage}%
  \hfill
  \vspace{4pt}
}

\def\paperTitle{\includegraphics[height=14pt]{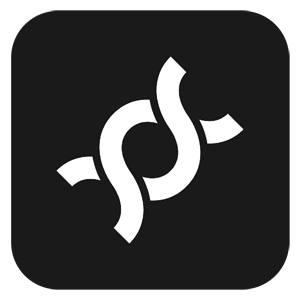} Evolver: Turning Agent Exploration into Evolving Capabilities}

\title{\paperTitle}

\def\authorBlock{
    Haoyang Zhang\footnotemark[2] \qquad Qumeng Sun \qquad Xiao Zhang \qquad  Xiang Liu\\
    \textbf{Zhidong Fan \qquad Junjie Wang}
    \\\\\fontsize{10}{10}
    \selectfont{Infinite Evolution Lab, EvoMap}
    \\\\\fontsize{10}{10}
    \url{https://github.com/EvoMap/evolver}
}

\author{\authorBlock}

\begin{document}
\evomapheader
\vspace{-0.08in}
\maketitle

{
  \renewcommand{\thefootnote}%
  {\fnsymbol{footnote}}
  \footnotetext[2]{Corresponding Author. {\tt\small 17@evomap.ai}}
}

\begin{abstract}
Agent exploration produces valuable procedural knowledge, yet much of this experience remains confined to transient execution traces and is not systematically reused.
We introduce \evolver, an experience-driven framework that transforms agent exploration into persistent, reusable capabilities without updating foundation model parameters.
Built around structured \gene assets, \evolver integrates experience acquisition, retrieval, and feedback-guided evolution to support experience-conditioned execution across existing agent harnesses.
We evaluate \evolver using Claude Code and Codex, two foundation model configurations, and four benchmarks: Terminal-Bench 2.1, SWE-bench Multilingual, SWE-rebench, and GAIA.
On a selected diagnostic set of 320 paired executions, \evolver improves task success from 83.13\% to 87.50\% (+4.38 percentage points) while reducing recorded task-execution token consumption by 24.24\%.
Furthermore, among 265 pairs successful under both conditions, token consumption decreases by 17.57\%, demonstrating efficiency benefits even when task success is preserved.
These findings suggest that persistent procedural experience can improve both task effectiveness and execution efficiency, providing a practical foundation for transforming agent exploration into evolving capabilities.
\end{abstract}

\section{Introduction}
\label{sec:introduction}

Large language model (LLM) agents have progressed from language generation to multi-step problem solving through tool use and environmental interaction, with growing applications in software engineering and other complex tasks~\cite{llm_agent_survey,software_agent_survey}.
Executing these tasks often requires extensive environment exploration, tool invocation, failure recovery, and outcome verification, incurring substantial computational and interaction costs.
Yet the procedural knowledge acquired through these interactions is often confined to individual executions, leaving subsequent tasks to rediscover useful strategies and repeat similar exploratory efforts.
This raises a fundamental question: \textit{How can the computational effort invested in past exploration be transformed into reusable experience that continuously improves future agent capabilities?}

Recent coding agents, such as Codex~\cite{openai_codex} and Claude Code~\cite{anthropic_claude_code}, have demonstrated the growing potential of autonomous software development.
Meanwhile, \skill-based approaches enable agents to reuse packaged instructions, workflows, and domain-specific resources~\cite{anthropic_agent_skills}, while other methods leverage historical trajectories, execution feedback, and evolving contexts to improve agent behavior without updating model parameters~\cite{expel,gepa,ace}.
More recently, research on strategy genes has shown that compact, control-oriented \gene representations can provide more effective experience reuse and iterative refinement than documentation-oriented \skill representations in controlled scientific coding settings~\cite{strategy_genes}.
Despite these advances, a central systems challenge remains: how to transform task-specific exploration into persistent, reusable strategy assets that can be integrated with existing agents and progressively refined through execution feedback.
Moreover, accumulating experience does not necessarily improve performance, as irrelevant or outdated strategies may introduce redundant computation or interfere with otherwise successful executions.

To address this challenge, we introduce \evolver, an experience-driven agent evolution system built around the principle of \textit{Exploration-to-Evolution}: transforming knowledge acquired through task execution into persistent strategy assets that can guide subsequent decisions and support iterative adaptation.
Building on the \gene representation, \evolver encodes reusable procedures and their applicability conditions as structured \gene objects, associates them with concrete execution cases through Capsules, and maintains event, provenance, and review records to support experience management.
Through integration with existing agent harnesses, \evolver makes historical experience available during subsequent task execution without modifying the underlying model parameters.
Rather than treating execution histories as passive records, \evolver organizes procedural knowledge into reusable assets that can be selected, applied, and maintained in response to new execution feedback.
This design establishes a mechanism through which previously invested exploration can contribute to future agent behavior, providing a foundation for experience-driven capability evolution.

We evaluate \evolver across two agent harnesses, Claude Code~\cite{anthropic_claude_code} and Codex~\cite{openai_codex}, two foundation model configurations, and four benchmarks: Terminal-Bench 2.1~\cite{terminal_bench}, SWE-bench Multilingual~\cite{swebench_multilingual}, SWE-rebench~\cite{swe_rebench}, and GAIA~\cite{gaia}.
On a selected diagnostic set of 320 paired executions, \gene augmentation improves the overall task success rate from 83.13\% to 87.50\% (+4.38 percentage points), with gains observed under both Claude Code (+2.50 points) and Codex (+6.25 points).
These improvements vary across models and benchmarks, with nine of the 16 evaluated configurations exhibiting higher task success.
Beyond task effectiveness, \evolver reduces recorded task-execution token consumption by 24.24\% overall.
Notably, among 265 pairs that succeed both with and without \evolver, token consumption decreases by 17.57\%, indicating that historical experience can improve execution efficiency even when task success is already achieved.
Representative cases in terminal operations and software repair further illustrate these efficiency gains through fewer model invocations and lower token usage without compromising task success.
Together, these findings highlight the complementary benefits of reusable procedural experience for task effectiveness and execution efficiency, positioning \evolver as a practical systems approach to transforming agent exploration into persistent, reusable capabilities without updating foundation model parameters.

\section{Related Work}
\label{sec:related_work}

\subsection{Procedural Experience Representation and Reuse}
\label{sec:rw_experience}

Agent interactions generate procedural knowledge through exploration, tool use, and execution feedback, motivating approaches that reuse historical experience without updating model parameters.
Early methods extract reusable insights from successful and failed trajectories~\cite{expel,reflexion}, while subsequent studies organize such knowledge into workflows, adaptive memory, and procedural guidance~\cite{agent_workflow_memory,dynamic_cheatsheet,lifemem,reme}.
Complementary approaches package reusable procedures as executable skills or structured instructions~\cite{voyager,agent_skills,skill_pro}, whereas Agentic Plan Caching~\cite{agentic_plan_caching} reuses adaptable planning templates to reduce repeated computation.
Recent studies further highlight that experience utility depends on representation quality and contextual applicability, as inappropriate skills may increase execution costs or degrade performance~\cite{skill_usage_wild,agent_skills_harmful}.

More closely related, the study~\cite{strategy_genes} demonstrates the advantages of compact, control-oriented \gene representations over documentation-oriented \skill packages for experience reuse and iterative refinement.
Building on this representational foundation, \evolver operationalizes procedural knowledge as persistent strategy assets associated with execution cases, provenance, and lifecycle states, enabling experience reuse within existing agent systems.

\subsection{Experience-Driven Agent Adaptation and Evolution}
\label{sec:rw_evolution}

Beyond experience reuse, recent research investigates how execution feedback can progressively improve agent behavior.
GEPA~\cite{gepa} and Agentic Context Engineering~\cite{ace} evolve prompts and contextual guidance through reflection and feedback, while EvolveR~\cite{evolver_rl} combines experience-derived principles with reinforcement learning.
Other approaches support continual adaptation through procedural memory refinement, skill evolution, and experience governance~\cite{skillrl,muse_autoskill,msce,memevolve}.
At the implementation level, Godel Agent~\cite{godel_agent} and Darwin Godel Machine~\cite{darwin_godel_machine} pursue self-improvement through modifications to agent code.

These studies demonstrate complementary pathways for agent evolution, spanning prompts, memory, skills, model policies, and executable programs.
In contrast, \evolver focuses on evolving external procedural experience while preserving the underlying foundation model and existing agent harnesses.
Its \textit{Exploration-to-Evolution} design connects experience acquisition, structured strategy reuse, and feedback-supported lifecycle management, allowing historical exploration to inform subsequent execution.
We examine the conditional benefits and computational trade-offs of experience reuse, recognizing that accumulated experience does not necessarily yield consistent improvements.

\section{\evolver: From Exploration to Evolution}
\label{sec:evolver}

\subsection{Design Principles and System Overview}
\label{sec:overview}

\begin{figure*}[!t]
\centering
\includegraphics[width=\textwidth]{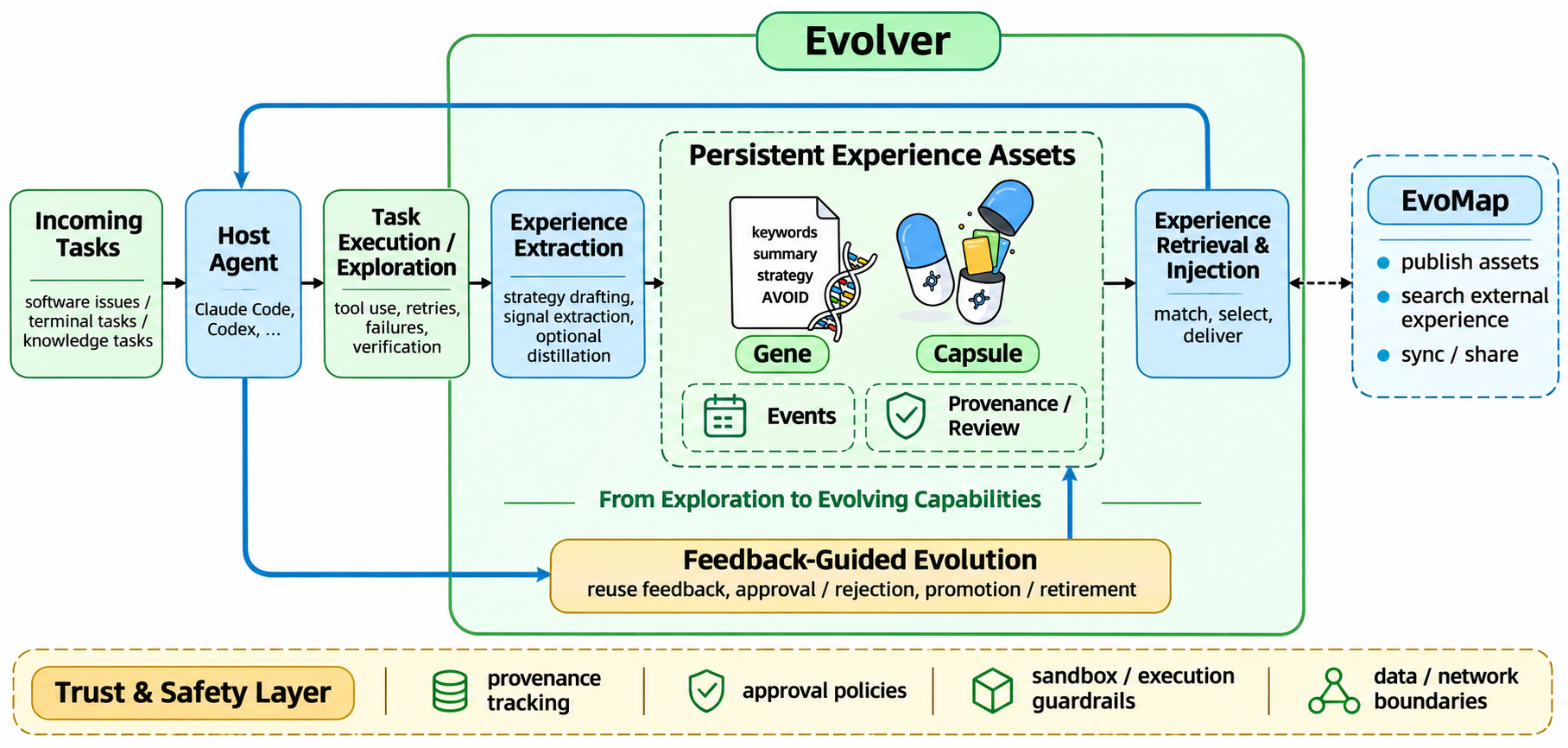}
\caption{
Overview of \evolver and its Exploration-to-Evolution framework, illustrating experience acquisition, persistent strategy assets, experience-conditioned execution, feedback-supported evolution, EvoMap connectivity, and operational trust boundaries.
}
\label{fig:overview}
\end{figure*}

\nbf{Problem Formulation.}
Let $\pi_\theta$ denote a fixed foundation-model policy, $h$ an existing agent harness, and $x_t$ the task encountered at step $t$.
Without updating model parameters, \evolver maintains an external experience state $\mathcal{E}_{t}$ to inform subsequent execution.
The interaction is formulated as
\begin{equation}
\begin{aligned}
g_t &= \mathcal{R}_h(x_t,\mathcal{E}_t),\\
\tau_t &= \operatorname{Exec}_{\pi_\theta,h}(x_t;g_t),\\
\mathcal{E}_{t+1} &= \mathcal{U}(\mathcal{E}_t,\tau_t,f_t),
\end{aligned}
\label{eq:evolver_cycle}
\end{equation}
where $\mathcal{R}_h$ retrieves applicable experience, $\tau_t$ denotes the resulting execution trajectory, and $f_t$ represents available execution feedback.
The update operator $\mathcal{U}$ abstracts experience acquisition and lifecycle operations, which depend on the enabled system components.

\nbf{Design Principles.}
\evolver follows the \textit{Exploration-to-Evolution} principle, which transforms knowledge acquired through agent exploration into persistent strategy assets that can support future adaptation.
This principle is realized through three complementary designs.
First, \textit{Experience Externalization} separates reusable procedural knowledge from individual execution trajectories, enabling its persistent representation and management.
Second, \textit{Experience-Conditioned Execution} makes relevant strategies available to existing agents without modifying their underlying model parameters or replacing their execution harnesses.
Third, \textit{Feedback-Guided Evolution} supports the assessment and maintenance of accumulated experience through execution feedback and lifecycle operations.
Together, these designs establish a pathway for preserving the knowledge value of historical exploration and making it available to subsequent agent executions.

\nbf{System Architecture.}
As illustrated in \cref{fig:overview}, \evolver integrates experience acquisition, persistent asset management, retrieval, and feedback-supported evolution into a unified system.
Historical agent interactions provide execution traces from which reusable strategies are extracted through signal analysis, automatic drafting, or optional model-based distillation.
The resulting experience is organized into structured \gene objects, associated \texttt{Capsule} records, and supporting event, provenance, and review information.
During subsequent tasks, relevant experience is retrieved and delivered through hooks, MCP interfaces, or host-specific integrations to condition agent behavior.
\evolver further connects to EvoMap~\footnote{\url{https://evomap.ai/}} for network-based experience publishing, discovery, and retrieval.
Provenance tracking, review mechanisms, and execution safeguards provide operational controls whose enforcement depends on the integration pathway and deployment environment.
The following sections detail these components and their roles in transforming historical exploration into reusable and evolving agent capabilities.

\subsection{Experience Representation and Asset Management}
\label{sec:experience_representation}

\nbf{Structured Strategy Representation.}
To preserve the knowledge acquired through historical exploration, \evolver represents reusable procedural knowledge as structured strategy assets rather than retaining complete execution trajectories as the primary unit of reuse.
Building on the strategy-gene representation~\cite{strategy_genes}, each \gene $g_i$ is characterized by
\begin{equation}
g_i = (s_i, \mathcal{S}_i, \mathcal{P}_i, \mathcal{M}_i, \mathcal{V}_i),
\label{eq:gene_representation}
\end{equation}
where $s_i$ summarizes the experience, $\mathcal{S}_i$ specifies procedural steps, $\mathcal{P}_i$ describes applicability conditions, $\mathcal{M}_i$ provides task-matching signals, and $\mathcal{V}_i$ contains optional validation plans.
This representation separates reusable operational guidance from task-specific execution details, allowing historical strategies to be retrieved and adapted to subsequent tasks.
Importantly, applicability conditions and validation plans provide guidance rather than enforceable execution constraints or guarantees of prior verification.

\nbf{Complementary Experience Assets.}
Beyond \gene objects, \evolver maintains complementary assets that preserve execution context and support experience management.
A \texttt{Capsule} associates a strategy with a concrete execution case, providing contextual examples that can accompany retrieved strategies.
An \texttt{AntiGene} captures negative procedural guidance, identifying actions or practices to avoid in subsequent executions.
Meanwhile, \texttt{Event} records capture asset-related operations, while separate provenance and review records track experience origins, trust status, and approval decisions.
Together, these components constitute the persistent experience state $\mathcal{E}_t$, separating reusable strategies, supporting execution cases, and operational metadata.

\nbf{Persistent Asset Management.}
\evolver maintains experience assets through content-addressed, append-only storage, preserving asset identity independently of subsequent lifecycle decisions.
Review and provenance states are maintained in separate records, supporting experience quarantine, approval, rejection, and revocation without overwriting the underlying assets.
This separation enables persistent procedural knowledge to coexist with evolving eligibility and trust states, providing a foundation for feedback-driven experience management.
The enforcement of these states depends on the corresponding acquisition and retrieval pathways, as detailed in~\cref{sec:experience_acquisition,sec:experience_retrieval,sec:experience_evolution}.

\subsection{Experience Acquisition and Consolidation}
\label{sec:experience_acquisition}

\nbf{Experience Sources.}
\evolver treats historical agent interactions as a source of reusable procedural knowledge, capturing not only task outcomes but also the exploration and feedback that precede them.
When session ingestion is enabled, runtime adapters collect records from supported agent harnesses, including task context, tool interactions, execution errors, and available outcomes.
These records are normalized into bounded experience materials with heuristic sensitive-information redaction.
Additional experience can be acquired through explicit asset submissions and external experience repositories, providing complementary sources beyond automatically collected sessions.

\nbf{Experience Extraction.}
Given collected experience materials, \evolver identifies lightweight signals associated with successful execution, failures, and task difficulty to construct candidate \gene objects.
The default automated drafting pathway extracts concise procedural hints from assistant utterances and marks them as unproven, rather than assuming that the resulting strategies have been validated.
When enabled, LLM-based distillation further supports structured experience extraction from execution records, including reusable strategies, associated cases, and negative procedural guidance.
These complementary pathways separate low-cost experience collection from more structured synthesis, while leaving the effectiveness of generated strategies subject to subsequent evaluation.

\nbf{Experience Consolidation.}
Extracted experience is consolidated into persistent, content-addressed assets, with supporting execution cases retained when available.
Automatically drafted and model-distilled candidates are placed under review, whereas certain explicit submission pathways can persist assets directly.
Review and provenance records maintain experience eligibility and source information separately from the underlying strategy content, although admission and trust enforcement depend on the acquisition and retrieval pathways.
By preserving candidate knowledge independently of its current lifecycle state, \evolver establishes a reusable experience base that supports subsequent retrieval and feedback-guided evolution, as detailed in~\cref{sec:experience_retrieval,sec:experience_evolution}.

\subsection{Experience Retrieval and Agent Integration}
\label{sec:experience_retrieval}

\nbf{Task-Conditioned Experience Retrieval.}
To reuse historical knowledge, \evolver provides multiple retrieval pathways that select experience according to task context and integration requirements.
Session-start hooks supply a bounded set of locally eligible \gene objects, while optional prompt-time recall matches task descriptions against stored experience signals.
MCP interfaces additionally support on-demand experience retrieval, whereas autonomous execution employs a separate heuristic selection mechanism incorporating task relevance, experience quality, and historical reuse feedback.
These pathways differ in their ranking strategies, context budgets, and eligibility checks, allowing experience delivery to accommodate different execution settings.

\nbf{Experience-Conditioned Execution.}
Retrieved \gene objects, together with associated \texttt{Capsule} examples when available, are delivered through session hooks, prompt-time injection, MCP tools, or host-specific integrations.
Rather than modifying foundation model parameters or replacing agent execution policies, \evolver introduces procedural guidance into the existing interaction context, enabling the host agent to interpret and apply historical strategies.
This separation between experience management and task execution allows \evolver to augment existing agent harnesses while preserving their native reasoning, tool-use, and verification capabilities.
Consequently, historical exploration can inform subsequent decisions without requiring additional model training or modifications to the underlying agent architecture.

\nbf{Networked Experience through EvoMap.}
Beyond local experience reuse, \evolver integrates with EvoMap to support the publication, discovery, and retrieval of experience assets across connected agents.
This interface extends the available knowledge beyond individual execution histories, allowing agents to access strategies contributed by other nodes.
External assets can be associated with provenance and trust metadata, although direct retrieval and locally imported assets follow different admission and review pathways.
Together, local and network-based experience access provides a foundation for broader experience reuse.

\subsection{Feedback-Guided Experience Evolution}
\label{sec:experience_evolution}

\nbf{Execution Feedback.}
Beyond preserving and reusing historical knowledge, \evolver supports experience evolution through feedback collected from subsequent executions.
Depending on the execution pathway, feedback includes task outcomes, validation results, reuse events, and interaction signals from agent hosts.
Autonomous execution evaluates task outcomes using execution status and available verification procedures, while host integrations can report observed checks and reuse-related signals.
These observations provide a basis for experience assessment.

\nbf{Experience Lifecycle.}
Rather than treating accumulated experience as permanently applicable, \evolver maintains lifecycle states that govern how strategies are retained and reused.
Automatically drafted and model-distilled \gene candidates initially enter quarantine and can subsequently be approved or rejected through review.
An optional probation mechanism further supports promotion based on recorded successes and failures, while autonomous execution incorporates historical reuse signals and failure-based restrictions into candidate selection.
Rejection and provenance revocation are recorded separately from the underlying content-addressed assets, preserving their identity while modifying their subsequent eligibility.
These mechanisms instantiate the lifecycle operations underlying $\mathcal{U}$ in~\cref{eq:evolver_cycle}, allowing experience states to evolve without modifying foundation model parameters.

\nbf{Trust-Aware Evolution.}
Because experience may originate from local execution or external contributors, \evolver associates assets with provenance and review information to support source-aware management.
Imported EvoMap assets can be marked as untrusted until explicitly promoted, while supported execution pathways apply additional validation and operational safeguards.
However, admission checks and trust enforcement are pathway-dependent, and some direct retrieval interfaces do not apply the same restrictions.
Together, these mechanisms provide a foundation for feedback-guided experience adaptation, distinguishing the evolution of reusable strategy states from an assumption of uniformly improving agent performance.

\section{Experimental Evaluation and Analysis}
\label{sec:experiments}

\subsection{Experimental Setup}
\label{sec:experimental_setup}

\nbf{Benchmarks and Agent Configurations.}
We evaluate \evolver on four benchmarks covering diverse agent capabilities: Terminal-Bench 2.1~\cite{terminal_bench}, SWE-bench Multilingual~\cite{swebench_multilingual}, SWE-rebench~\cite{swe_rebench}, and GAIA~\cite{gaia}.
Terminal-Bench 2.1 evaluates terminal-based engineering tasks, SWE-bench Multilingual and SWE-rebench assess software issue resolution, and GAIA examines multi-step information seeking and reasoning.
We consider two agent harnesses, Claude Code and Codex, each paired with two foundation model configurations, \texttt{deepseek-v4.1-flash}~\cite{deepseek_v41_flash} and \texttt{gpt-5.6-sol}~\cite{gpt_56_sol}.
For each benchmark, we prioritize the first 20 tasks in their original order.
Tasks yielding empty responses due to malformed outputs, timeouts, or execution failures are replaced sequentially with subsequent tasks.
We ensure that all harness--model configurations are evaluated on the same 20 tasks per benchmark, yielding 16 configurations in total.
All experiments use \evolver v2.0.33.

\nbf{Experience Construction and Paired Evaluation.}
We adopt a three-stage \textit{Baseline--Extraction--Gene} (B--E--G) protocol to examine the effects of historical experience reuse.
In the \textit{Baseline Execution} (B) stage, each agent executes the original task using its native tools and built-in skills without additional \gene injection.
In the \textit{Experience Extraction} (E) stage, reusable strategies are distilled from valid baseline trajectories, including both successful and unsuccessful executions, using a model configuration matched to the corresponding evaluation setting.
Experience pools are isolated by evaluation batch, benchmark, harness, and model, and candidate assets are frozen before augmented execution.
In the \textit{Gene-Augmented Execution} (G) stage, the agent receives the original task together with selected \gene guidance in a fresh session with the task environment restored.
Gene selection prioritizes experience derived from the same task, followed by other matching candidates within the isolated pool, with an injection budget of 12,000 bytes.
Both B and G retain the same foundation model and harness configuration and are evaluated using the corresponding native benchmark grader.
This protocol assesses historical experience reuse rather than generalization to unseen tasks or long-term online evolution.

\nbf{Evaluation Criteria and Metrics.}
We analyze 320 paired observations, retaining 20 tasks per configuration with valid outcomes and complete token records for both B and G, excluding executions interrupted by timeouts or token limits.
The final task selection additionally considers observed outcome improvements and token usage, making the resulting set diagnostic rather than a random sample of the full benchmarks.
We report task success rates and percentage-point differences, together with recorded task-execution tokens and model invocation counts.
Aggregate token reduction is computed from total usage across paired executions rather than averaging individual reduction ratios.
We further analyze pairs that succeed under both conditions to assess resource changes while preserving task success.
Token accounting includes recorded model input and output usage but excludes experience extraction and other overhead outside the final task executions.

\subsection{Overall Task Performance}
\label{sec:overall_results}

\begin{figure*}[ht]
    \centering
    \includegraphics[width=0.82\textwidth]{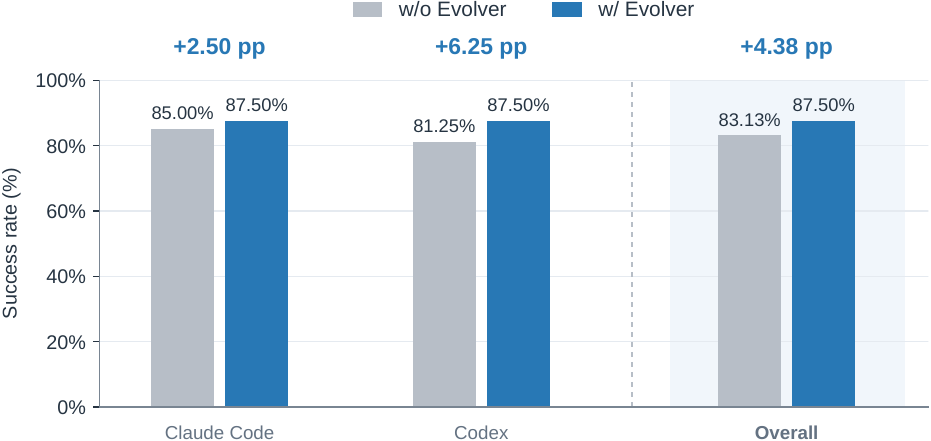}
    \caption{
    Overall task success with and without \evolver across two agent harnesses and the combined evaluation set.
    }
    \label{fig:overall_performance}
\end{figure*}

\nbf{Overall Effectiveness.}
As shown in~\cref{fig:overall_performance}, incorporating historical \gene experience increases the overall success rate from 83.13\% to 87.50\%, an improvement of 4.38 percentage points across 320 paired evaluations.
These results demonstrate the potential of reusable procedural experience to improve task completion within the selected evaluation setting.

\nbf{Performance across Agent Harnesses.}
Both evaluated harnesses exhibit improvements when augmented with \evolver.
Claude Code improves from 85.00\% to 87.50\% (+2.50 percentage points), while Codex increases from 81.25\% to 87.50\% (+6.25 percentage points).
These consistent gains across both harnesses highlight the applicability of experience-conditioned execution to different agent runtimes without modifying their underlying model parameters.

\subsection{Performance across Models and Benchmarks}
\label{sec:detailed_results}

\begin{figure*}[h]
    \centering
    \includegraphics[width=0.88\textwidth]{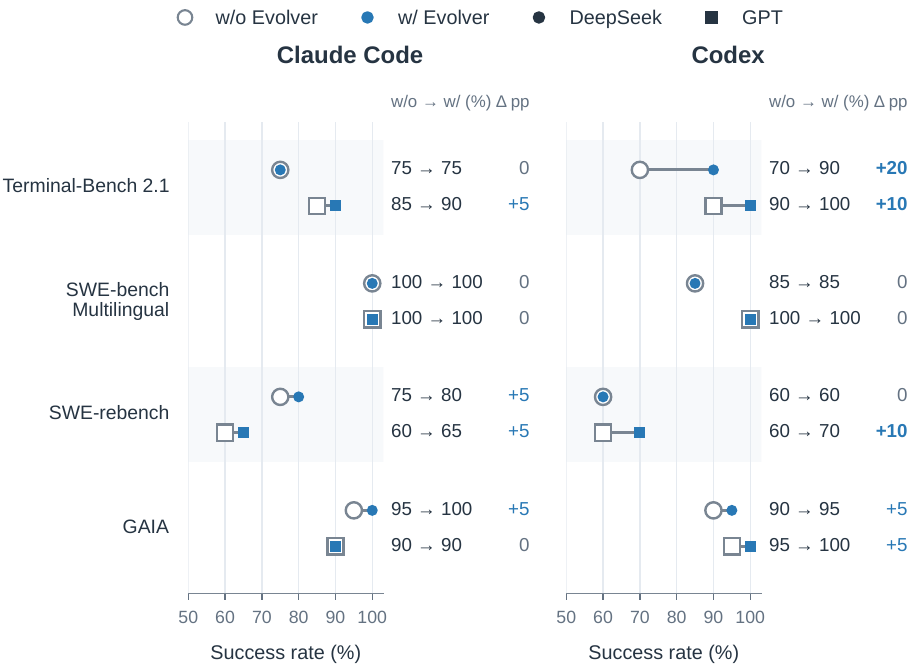}
    \caption{
    Task success across agent harnesses, foundation models, and benchmarks, with 20 paired evaluations per configuration.
    }
    \label{fig:configuration_performance}
\end{figure*}

\nbf{Model-Dependent Performance.}
As shown in~\cref{fig:configuration_performance}, the benefits of experience reuse vary across foundation models and agent harnesses.
Among the 16 evaluated configurations, nine exhibit improved task success, while seven remain unchanged.
Under Codex, DeepSeek achieves the largest improvement on Terminal-Bench 2.1, increasing from 70\% to 90\% (+20 percentage points), whereas GPT improves by 10 points on both Terminal-Bench 2.1 and SWE-rebench.
For Claude Code, both models improve by 5 points on SWE-rebench, while additional gains occur on different benchmarks: GAIA for DeepSeek and Terminal-Bench 2.1 for GPT.
These results suggest that the effectiveness of historical procedural guidance depends on the underlying model and its interaction with the agent harness.

\nbf{Benchmark-Dependent Performance.}
Across the evaluated configurations, Terminal-Bench 2.1 exhibits the largest aggregate improvement, increasing from 64/80 to 71/80 (+8.75 percentage points).
SWE-rebench and GAIA also improve by 5.00 and 3.75 points, respectively, demonstrating benefits across software engineering and information-seeking tasks.
In contrast, SWE-bench Multilingual maintains the same success rate of 96.25\% under both conditions.
These findings highlight the task-dependent nature of experience reuse, while motivating further analysis of whether historical knowledge can reduce execution costs even when task success remains unchanged.

\subsection{Token Efficiency of Experience Reuse}
\label{sec:efficiency}

\begin{figure*}[h]
    \centering
    \includegraphics[width=0.80\textwidth]{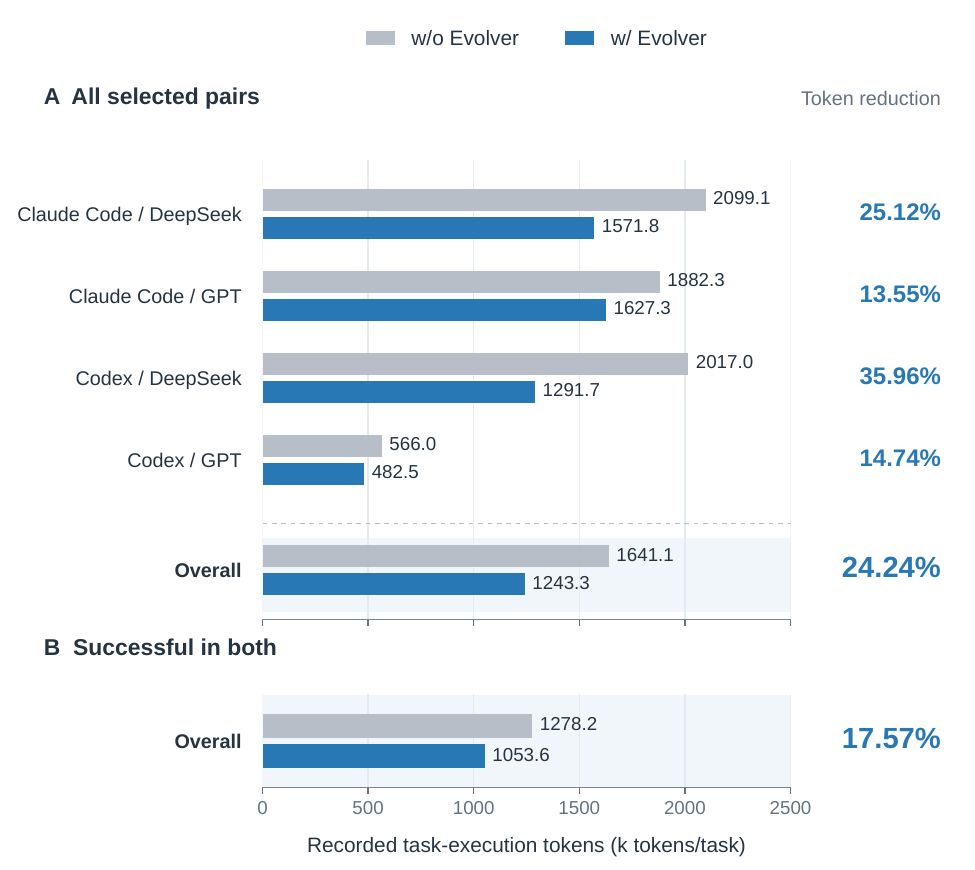}
    \caption{
    Recorded task-execution token usage with and without \evolver across all selected pairs (A) and pairs successful under both conditions (B).
    }
    \label{fig:token_efficiency}
\end{figure*}

\nbf{Overall Token Reduction.}
As shown in~\cref{fig:token_efficiency}(A), \evolver reduces average task-execution token consumption from 1,641.1K to 1,243.3K, corresponding to an aggregate reduction of 24.24\% across 320 paired evaluations.
Token usage decreases across all four harness--model configurations, with reductions ranging from 13.55\% to 35.96\%.
The largest reduction occurs with Codex and DeepSeek (35.96\%), followed by Claude Code and DeepSeek (25.12\%).
These results indicate that historical procedural experience can reduce the computational resources required for subsequent agent execution across different configurations.

\nbf{Efficiency under Preserved Success.}
We further consider \textit{Successful in Both}, a subset of 265 paired tasks that pass the native benchmark evaluation both with and without \evolver.
As shown in~\cref{fig:token_efficiency}(B), average token usage decreases from 1,278.2K to 1,053.6K, yielding a 17.57\% reduction while maintaining successful task completion.
This finding highlights the value of reusable experience beyond improving task success, demonstrating its potential to reduce execution overhead for tasks that agents can already solve.

\subsection{Case Studies}
\label{sec:case_studies}

\begin{figure*}[h]
    \centering
    \includegraphics[width=0.96\textwidth]{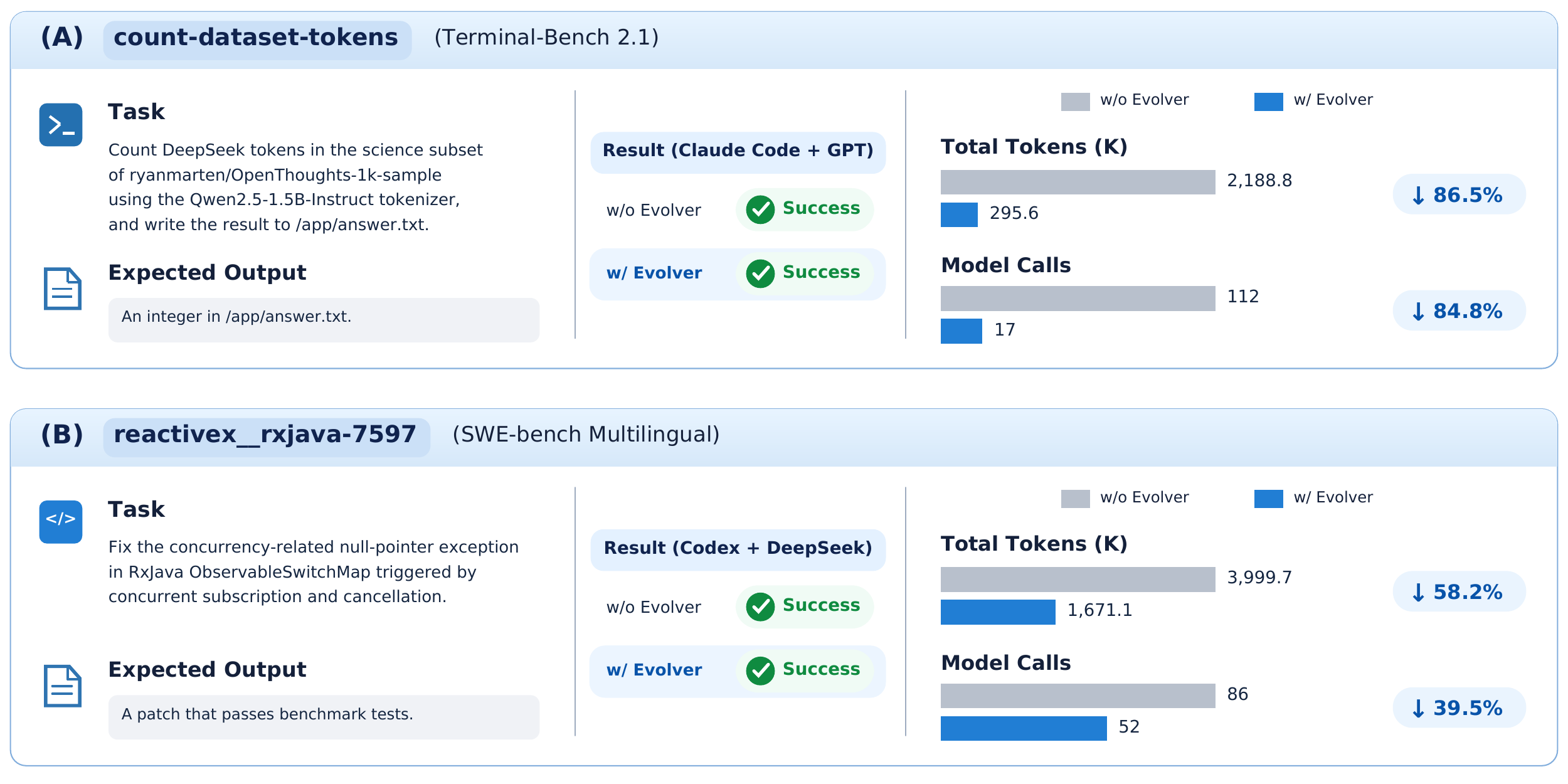}
    \caption{
    Representative cases showing preserved task success with reduced token consumption and model invocations.
    }
    \label{fig:case_studies}
\end{figure*}

To complement the aggregate results, \cref{fig:case_studies} presents two representative tasks where historical experience reduces execution resources without changing task success.

\nbf{Dataset Token Counting.}
As illustrated in \cref{fig:case_studies}(A), this Terminal-Bench 2.1 task requires counting tokens in a specified dataset subset using a designated tokenizer and producing an integer result.
With Claude Code and GPT, both executions succeed, while \evolver reduces token consumption by 86.5\% and model invocations from 112 to 17.

\nbf{Software Repair.}
As illustrated in \cref{fig:case_studies}(B), this SWE-bench Multilingual task involves resolving a concurrency-related null-pointer exception in RxJava's \texttt{ObservableSwitchMap}.
With Codex and DeepSeek, both executions pass the benchmark evaluation, while \evolver reduces token consumption by 58.2\% and model invocations from 86 to 52.

\section{Conclusions}
\label{sec:conclusions}

In this work, we introduced \evolver, a framework for transforming transient agent exploration into persistent, reusable procedural experience.
By integrating structured \gene representation, experience acquisition, retrieval, and feedback-guided evolution, \evolver enables existing agent harnesses to reuse accumulated knowledge without modifying foundation model parameters.
Across two agent harnesses, two foundation model configurations, and four benchmarks, our selected diagnostic evaluation shows that experience reuse improves task success by 4.38 percentage points while reducing recorded task-execution token consumption by 24.24\%.
Notably, token consumption decreases by 17.57\% among tasks successfully completed under both conditions, suggesting that reusable experience can improve execution efficiency even when task success remains unchanged.
These findings highlight the complementary roles of accumulated procedural knowledge in improving task effectiveness and reducing redundant execution effort.
Future work will investigate more reliable experience selection and validation, transfer to unseen tasks, and longitudinal evolution of reusable capabilities.
More broadly, \evolver provides a systems foundation for turning isolated agent interactions into persistent knowledge that can support continued capability improvement.

\bibliographystyle{unsrtnat}
\bibliography{references}

\end{document}